# Enhancement of $J_c$ by doping silver in grain boundaries of YBa₂Cu₃Oᵧ polycrystals with solid-state diffusion method

C. H. Cheng and Y. Zhao[a)]

*Superconductivity Research Group, School of Materials Science and Engineering,*
*University of New South Wales, Sydney 2032, NSW, Australia*

Grain boundaries (GBs) of YBa₂Cu₃Oᵧ (YBCO) are preferentially doped with Ag by the solid-state diffusion method. The distribution of Ag is highly localized in the GB. As a result of the Ag doping, the $J_c$-$H$ behavior of YBCO textured polycrystalline samples is significantly improved, and at 60 K and 7 T, $J_c$ is tripled. It is proposed that Ag could partially replace the Cu in CuO₂ planes near GBs, therefore reducing the geometric distortion of Cu-O bonds and making the effective GBs thinner. © 2003 American Institute of Physics. [DOI: 10.1063/1.1539909]

For practical applications of high temperature superconductors (HTSs), long wires/tapes with a proper thickness and a sufficient engineering critical current must be produced. To achieve this goal, the problem of grain boundaries (GBs), which drastically degrade the performance of the wire,[1] must be overcome. It is generally considered that the weak-link effect in GBs is one of the main obstacles reducing the $J_c$ of HTSs.[1–5] For instance, owing to their extremely anisotropic nature, textured Bi2223 tapes contain about 80% of [001] or c-axis twist GBs[4] forming so-called colonies.[5] The colonies are connected through tilt GBs. In textured tapes, the colonies are mostly connected through [100] or [010] or mixed off-plane tilt GBs. A significant superconductivity depression (around 10 K) is observed on the [100] tilt GBs[6] in the Bi system. In YBa₂Cu₃Oₓ (YBCO), it is also found that the critical current density decreases rapidly as the misorientation angle $\vartheta$ of the GBs increases.[2,3]

Many efforts have been made to develop the means to avoid the occurrence of high-angle GBs in HTS tapes/wires. Practically, it is very hard to avoid the occurrence of high-angle GBs in kilometer-long HTS tapes/wires. Therefore, a more realistic approach of realizing the large-scale application of HTSs is to "repair" these faulty parts and to pursue "weak-link-free" instead of "GB-free" conductors. As is reported recently,[7,8] by doping Ca at the Y site, the GB critical current density in the zero applied field is enhanced by about ten times. The GB critical current density of the thin-film bicrystals of YBCO in a high field is also improved by doping Ca; for example, the GB critical current density at 44 K is increased by as high as 30% over 0–3 T and threefold at 5 T.[9] Similar results are obtained in Ca-doped YBCO textured polycrystalline bulks.[10] All these demonstrate that the "damage" of superconductivity in GBs of HTSs can be "repaired" to a large extent by chemical doping.

On the other hand, Ag has been widely used in the fabrication of HTS wires/tapes, and as a dopant or additive to improve the microstructure, and thus the $J_c$ of HTSs. Up to now, the understanding of the role of Ag in HTSs is limited to that of Ag being used to fill microcracks or pores in the microstructure. Usually, as the microcracks or pores in HTSs are much larger in size than the coherence lengths of HTSs, Ag doping could not significantly improve the $J_c$ if Ag works only as filler. However, it is highly possible that Ag plays a much bigger role in HTSs. As is well known, Bi2223 has a larger anisotropy and a smaller coherence length than YBCO, and thus should have a more drastic weak-link effect. Contrary to this, the weak-link effect is more drastic in YBCO than in Ag-clad Bi2223 tapes. Therefore, it should be asked: Besides being a metal sheath, does Ag play any other roles in Bi2223/Ag tapes? Another phenomenon to which we should pay attention is the tendency of the highest-local $J_c$ in Bi2223/Ag tapes often being near the Bi2223-Ag interface.[11] These two examples imply that Ag may play a more important role in improving the $J_c$ of HTSs.

In this article, the solid-state diffusion method is employed to preferentially dope Ag into the GBs of YBCO. The changes of $J_c$ induced by the Ag doping are examined. It is found that Ag mainly diffuses to the GBs of YBCO, and the $J_c$ is improved significantly. The results indicate that Ag has played a role to repair the GBs in HTSs.

Melt-textured polycrystals with a c-axis alignment were prepared using the modified melt-textured-growth (MTG) method.[12] The reason behind the use of polycrystals as samples in this study was to highlight the GBs effect. Presintered bars with typical dimensions of $40 \times 15 \times 4$ mm³ were prepared with high purity (99.99%) YBCO+wt 25% Y₂BaCuO₅ (Y211) powders. Y211 substrate was used to absorb the unreacted liquid phase during the MTG process. Regular shapes with dimensions of $1.2 \times 1.0 \times 0.3$ mm³ were cut from the as-grown crystals, annealed in flowing argon at 500 °C for 20 h to pump out the oxygen from GBs, then coated *in situ* with Ag by thermal evaporate. The coated samples underwent a solid-state diffusion reaction in flowing oxygen at 850 °C for 20, 40, 60, 80, 100, and 120 h, respectively. For the samples which underwent a solid-state diffusion of less than 120 h, they were annealed, before being coated with Ag, at 850 °C in flowing oxygen for an appropriate duration in order to make the total annealing-time in oxygen of 120 h. After the diffusion reaction, the samples were polished and a layer of 20 $\mu$m was removed. Two pieces of crystals without Ag coating were annealed in the

a)Author to whom all correspondence should be addressed; electronic mail: y.zhao@unsw.edu.au

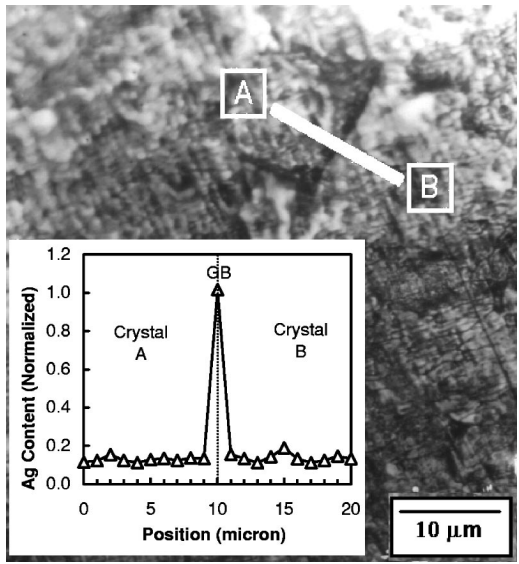

FIG. 1. Micrographs for the typical textured YBCO polycrystals doped with Ag. Inset: EPMA analysis result taken along the straight line A-B marked in Fig. 1.

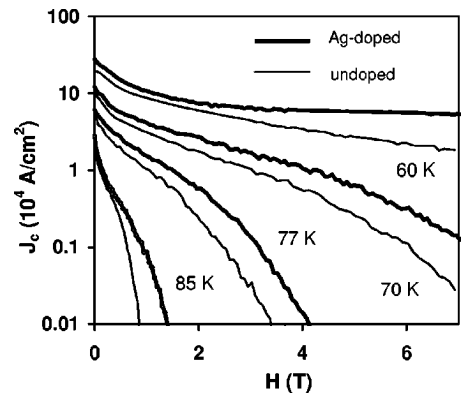

FIG. 2. Field dependence of $J_c$ for Ag-doped (reaction time is 100 h) and undoped YBCO textured polycrystals.

same conditions and used as references. X-ray diffraction revealed that all of these samples were well $c$-axis aligned, with no detectable differences in their crystal structures. The microstructure and composition of the samples were determined by scanning electron microscope and electron probe microanalyses (EPMA). A dc superconducting quantum interference device magnetometer was used to measure the magnetization loops (with $H//c$). $J_c$ was deduced from the $M(H)$ loops by the Bean model. $T_c$ values of these samples were almost the same (with a difference of less than 0.3 K), approximately equaling to 93.2 K.

Figure 1 shows a typical micrograph of the polished surfaces of our samples. As they are polycrystals with their $c$-axis aligned, many [001] tilt grain boundaries can be seen on their surfaces. The distribution of the dopant, as determined with EPMA, is highly localized around GBs. The inset of Fig. 1 shows the typical results of the EPMA analyses taken along the line "A-B" in Fig. 1. It is evident that the concentrations of Ag in GBs are about ten times higher than in the grains, indicating that Ag mainly stays in the GBs of YBCO. Such a preferential doping effect in GBs may be explained by the existence of the so-called "GB energy." As point defects, chemical dopants may occupy various positions in a real crystal, forming substitutional or interstitial impurities. As GBs are structurally distorted regions in crystals, extra energy forms in the GB regions due to the distortion. For small angle GBs, the GB energy $E_{GB}$ increases with increasing misorientation angle $\vartheta$. For large angle GBs, $E_{GB}$ is almost independent of $\vartheta$. Generally speaking, the relationship between $E_{GB}$ and $\vartheta$ can be described as[13]

$$E_{GB}=E_{GB}^0 \vartheta(A-\ln \vartheta),\qquad(1)$$

where $E_{GB}^0$ and $A$ are constants. Due to the existence of GB energy as well as the Coulomb interaction between the GB and the impurity atoms, the GB tends to attract impurity atoms in order to decrease the GB energy. Therefore, the chemical dopant has a higher probability of staying in the GB region than staying inside the crystal by a factor $C$. This factor is temperature dependent and obeys the following expression:

$$C=\exp(\Delta E/k_B T),\qquad(2)$$

where $\Delta E$ is the energy difference between the chemical dopant staying inside the crystal and the chemical dopant staying in the GB, $k_B$ is the Boltzmann constant, and $T$ is the temperature. This equation provides a theoretical basis for the preferential occupation of a chemical dopant in GBs. From the distribution of the dopant concentration in both the grains and the GB shown in the inset of Fig. 1, the energy difference $\Delta E$ between an Ag atom staying inside the crystal and an Ag atom staying in the GB in YBCO can be estimated, using Eq. (2), to be $\Delta E \sim 0.22$ eV (where $T=850\,°C$).

Figure 2 shows the typical magnetic-field dependence of $J_c$ for undoped and Ag-doped YBCO textured polycrystals at various temperatures between 60 and 85 K. The reaction time for this Ag-doped sample is 100 h. For the undoped sample, the $J_c$ value is $2.3\times10^4$ A/cm$^2$ at 85 K in the self field, $5.3\times10^4$ A/cm$^2$ at 77 K in the self field, and $1.0\times10^4$ A/cm$^2$ at 77 K in 1 T. Such a high $J_c$ value at 77 K indicates that the GBs in these textured polycrystalline samples are rather strongly coupled. However, the $J_c$ decreases rapidly with an increasing applied magnetic field. This may be attributed to the degradation of $J_c$ in the GB region.

A significant improvement in the $J_c$-$H$ behavior can be clearly seen from the results of the Ag-doped sample at all temperatures examined in this study, i.e., between 60 and 85 K. The improvement is reflected by an increase of $J_c$ in both low and high fields with the increase in the high fields being more significant. For example, at 77 K, $J_c$ is increased slightly in the zero applied field, but is nearly doubled in 1 T. At 60 K and 7 T, $J_c$ is nearly tripled. The improvement of $J_c$ is also dependent on the diffusion time of the sample reacting with Ag. As shown in Fig. 3, when the diffusion time is less than 20 h, the improvement is not very significant. When the reaction time is between 20 and 60 h, the improvement of $J_c$ becomes very remarkable. Further increasing the diffusion time results in a saturation of $J_c$.

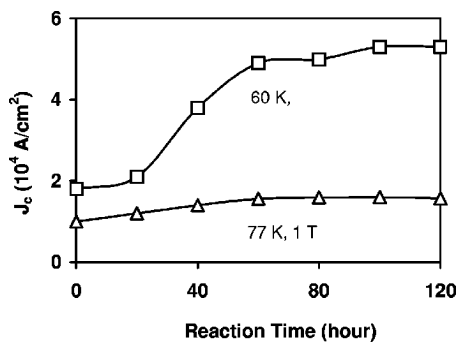

FIG. 3. Dependence of $J_c$ on the duration time of the diffusion reaction between Ag and YBCO.

Differing from the Ca doping that replaces $Y^{3+}$ with $Ca^{2+}$ and thus increasing the hole concentration of YBCO, Ag usually cannot be doped into the crystal structure of YBCO in bulk crystals. In order to understand the Ag doping effect on the improvement of the GB $J_c$, a model of the Ag-doping in GBs of HTSs is hereby proposed. Before presenting this model, it is necessary to examine the arrangement of atoms in the $CuO_2$ plane of a high-angle [001] GB. As shown in Fig. 4, Cu–O bonds near a high-angle [001] GB are significantly distorted or even broken by various misarrangements of the atoms. For example, the oxygen atom between the Cu atoms marked with "4" is squeezed away owing to a too-small spacing between these two Cu atoms; on the other hand, the Cu–O bonds along 1-2-1 and 6-5-6 are stretched owing to a too-big spacing between the atoms. The distortion (shortening and stretching) of the Cu–O bonds is not localized in one or two atomic layers near the GB, but is extended to a larger range from the GB because of the thermal relaxation of the distortion[14] and hence results in a "distorted region" covering many atomic layers away from the GB. As the electronic state of the Cu–O bonds is very sensitive to this distortion,[14] the "distorted region" may have poor superconductivity or even no superconductivity, leading to a "weak-link" effect in the GBs. However, the Cu atoms in the stretched Cu–O bonds, such as the Cu in position 1 and position 5, have the potential to be replaced by Ag due to the larger ionic size of Ag. When this replacement takes place, the geometric distortion of the Cu–O bond at the GB can be repaired by forming an Ag–O bond at 1-2-1 sites or 6-5-6 sites. This can in turn reduce the distortions of other Cu–O bonds next to the Ag–O bond. It is quite possible that these Ag–O bonds are nonsuperconducting but, are geometrically localized in the GB. In other words, the replacement of Cu with Ag transforms an "extended geometric distortion" into a "localized electronic distortion," which may greatly reduce the thickness of the "distorted region" and thus weaken the weak-link effect. Therefore, according to the proposed model, the mechanism of repairing GBs with doping Ag is quite different from that with doping Ca. Ca-doping compensates the carrier loss in GBs, reducing the GB resistance and thus increasing the GB $J_c$, whereas Ag doping reduces the thickness of the effective GB, weakening the weak-link effect of the GB. In our model, Ag merely replaces the Cu in the stretched Cu–O bond near GBs, therefore, only a part of the Cu–O bonds can be repaired by doping Ag. Using the same idea, the GBs of HTS may be further repaired with other dopants or with the concurrent doping of two or more elements.

In summary, we have investigated the doping effect of Ag on the GB $J_c$ of YBCO by preferentially doping GBs through solid-state diffusion. We have found that the distribution of Ag is highly localized in the GBs of YBCO. The $J_c$-$H$ behavior of YBCO textured polycrystalline samples is significantly improved by doping Ag in GBs. A mechanism for repairing GBs by doping Ag is proposed, which suggests that Ag could partially replace the Cu in $CuO_2$ planes near GBs, reducing the geometric distortion of the Cu–O bonds and making the effective GBs thinner.

This work was supported in part by the Australian Research Council under Grant No. F89700792.

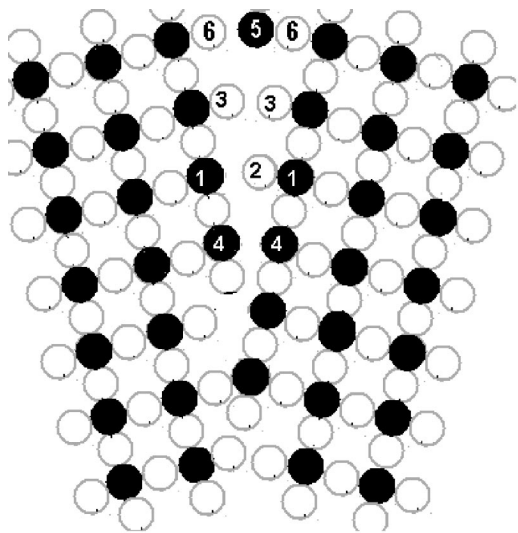

FIG. 4. A schematic of atom positions in the $CuO_2$ plane of a high angle [001] GB. The black balls represent the Cu atoms and the white represent the O atoms.

[1] A. Goyal et al., J. Mater. **51**, 19 (1999).
[2] D. Dimos, P. Chaudhari, and J. Mannhart, Phys. Rev. B **41**, 4038 (1990).
[3] D. Dimos, P. Chaudhari, J. Mannhart, and F. K. LeGoues, Phys. Rev. Lett. **61**, 219 (1988).
[4] Y. Zhu, M. Suenaga, and R. L. Sabatini, Appl. Phys. Lett. **65**, 1832 (1994).
[5] B. Hensel, G. Grasso, and R. Flukiger, Phys. Rev. B **51**, 15 456 (1995).
[6] Y. N. Tsay, Q. Li, Y. Zhu, M. Suenaga, K. Shibutani, I. Shigaki, and Ogawa, IEEE Trans. Appl. Supercond. **9**, 1622 (1999).
[7] A. Schmehl, B. Goetz, R. R. Schulz, C. W. Schbeider, H. Bielefeldt, H. Hilgenkamp, and J. Mannhart, Europhys. Lett. **47**, 110 (1999).
[8] G. Hammeri, A. Schmehl, R. R. Schulz, B. Goetz, H. Bielefeldt, C. W. Schbeider, H. Hilgenkamp, and J. Mannhart, Nature (London) **407**, 162 (2000).
[9] A. G. Daniels, A. Gurevich, and D. C. Larbalestier, Appl. Phys. Lett. **77**, 3251 (2000).
[10] C. H. Cheng and Y. Zhao, Supercond. Sci. Technol. **16**, 130 (2003).
[11] A. E. Pashitski, A. Polyanskii, A. Gurevich, J. A. Parrell, and D. C. Larbalestier, Physica C **246**, 133 (1995).
[12] C. H. Choi, Y. Zhao, C. C. Sorrell, M. La Robina, and C. Andrikidis, Physica C **269**, 306 (1996).
[13] B. K. Xu, W. P. Yan, and M. D. Liu, *Crystallography* (Press of Jilin University, Changchun, 1991), p. 264.
[14] S. V. Stolbov, M. K. Mironova, and K. Salama, Supercond. Sci. Technol. **12**, 1071 (1999).